\documentstyle[multicol,epsf,aps]{revtex}
\begin{document}
\title{Mean Field Theory of Polynuclear Surface Growth}
\author{E.~Ben-Naim$^1$, A.~R.~Bishop$^1$, I.~Daruka$^{1,2}$,
  and P.~L.~Krapivsky$^3$}
\address{$^1$Theoretical Division and Center for Nonlinear Studies, 
  Los Alamos National Laboratory, Los Alamos, NM 87545}
\address{$^2$Department of Physics, University of Notre Dame, Notre
  Dame, IN 46556}
\address{$^3$Center for Polymer Studies and Department of Physics,
  Boston University, Boston, MA 02215}
\maketitle 
\begin{abstract} 
  
  We study statistical properties of a continuum model of polynuclear
  surface growth on an infinite substrate.  We develop a
  self-consistent mean-field theory which is solved to deduce the
  growth velocity and the extremal behavior of the coverage.
  Numerical simulations show that this theory gives an improved
  approximation for the coverage compare to the previous linear
  recursion relations approach.  Furthermore, these two approximations
  provide useful upper and lower bounds for a number of
  characteristics including the coverage, growth velocity, and the
  roughness exponent.

\end{abstract}

\begin{multicols}{2}

\section{Introduction}

Kinetics of surface growth is a fascinating field that has been the
subject of intense current research\cite{god,meakin,zhang,barabasi}.
It is well established that as the surface grows its morphology
remains scale invariant, and for example, fluctuations in the
interface height exhibit an asymptotic scaling behavior. While the
understanding of growth on one-dimensional (1d) substrates is rather
comprehensive, in the physical case of two-dimensional substrates current
theoretical understanding remains incomplete\cite{zhang,barabasi}.

In this study we focus on an appealingly simple yet non-trivial
surface growth problem, the so-called polynuclear growth (PNG)
model\cite{frank,kas,gilmer}.  The PNG model describes the
evolution of islands that nucleate at random on top of previously
nucleated islands and grow in the radial direction. This model is
appropriate for describing situations where there is a competition
between growth along the step edges and growth due to nucleation, as
is the case in polymer crystal growth\cite{keller,bassett}.

The submonolayer version of the PNG model is identical to the
classical Kolmogorov-Avrami-Johnson-Mehl (KAJM) nucleation-and-growth
process\cite{Kolm,Avrami,Johnson} where exact results for the coverage
and more detailed statistical properties are possible
\cite{sek,Sek,bradley,ep}.  Using the exact solution for the KAJM
coverage, an approximate Linear Recursion Relations (LRR) approach to
the PNG process in arbitrary dimension was suggested
\cite{kas,Evans,be,new}.  Additionally, steady states on 1d substrates
were obtained analytically using the fact that kinks (antikinks) are
uncorrelated \cite{frank,ben,gold}.  However, the non-equilibrium
behavior and especially the asymptotic time dependence remains an open
problem, despite a number of studies
\cite{be,new,ben,gold,gil,krug,kr}.

Our goal is to develop a self-consistent approach that can be viewed
as a mean field theory (MFT) of the PNG process, and to compare it
with the previous LRR approximation as well as with numerical
simulations.  We will show that MFT offers a better description for
the time dependent coverage, and that the two approaches, when
combined, provide upper and lower bounds for the coverage and the
growth velocity.

The rest of this paper is organized as follows.  In the next section
we define the PNG model. In Sec. III, we develop a self-consistent
mean field approximation.  We find that possible growth velocities are
bounded from below, $v\geq v_{\rm min}$, and show that the minimal
velocity is actually selected.  We also solve for the coverage profile
in the tail regions.  In Sec.~IV, we review the LRR approach and
derive the coverage profile analytically.  In Sec.~V, we present
simulation results. Conclusions are given in Sec.~VI.

\section{THE PNG MODEL}

The PNG model is defined as follows.  Consider a flat uniform
$d$-dimensional substrate at time $t=0$.  Seeds of negligible size
nucleate randomly at a constant rate per unit area, and grow with a
constant velocity in the radial direction.  When two islands on the
same layer meet they coalesce, and the joint perimeter continues
growing in the corresponding radial direction.  Meanwhile, nucleation
continuously generates additional layers on top of previously
nucleated layers.  Clearly, there are no overhangs in this model, a
feature that considerably simplifies the analysis.  Another important
simplification in the PNG model is that the nucleation rate is uniform
in time as well as in space, i.e., it is independent of the local
surface structure. Without loss of generality, we set the nucleation
rate and the radial growth velocity to unity. This can be achieved by 
an appropriate rescaling of space and time.

In this study, we concentrate on $S_j(t)$, the uncovered fraction in
the $j$th layer at time $t$.  This important characteristic of
multilayer growth gives the net exposed fraction of the $j$th layer,
$S_j(t)-S_{j-1}(t)$, and therefore can be used to calculate relevant
statistical properties. In general $\langle f(j)\rangle =
\sum_{j=0}^{\infty} f(j)[S_{j+1}(t)-S_j(t)]$, and in particular the
average height is given by

\begin{equation}
\label{h}
h(t)=\langle j\rangle= \sum_{j=1}^\infty j\left[S_{j+1}(t)-S_j(t)\right].
\end{equation} 
Fluctuations in the height are quantified by the mean square width or
roughness, $w^2(t)$,
\begin{eqnarray}
\label{w}
w^2(t)&=&\langle j^2\rangle -\langle j\rangle^2\\
&=&\sum_{j=1}^\infty j^2 \left[S_{j+1}(t)-S_j(t)\right]-h^2(t).\nonumber
\end{eqnarray} 
We expect a linear growth in time for the average height, $h(t)\simeq
vt$, and an algebraic growth for the interface width, $w(t)\sim
t^{\beta}$, with a priori unknown roughness exponent $\beta$.  In other
words, the uncovered fraction obeys the following wave-like form
\begin{equation}
\label{form}
S_j(t)=F\left(j-vt\over t^{\beta}\right). 
\end{equation}
The argument reflects the overall shift in the position of the
wave-front, and the multiplicative scale accounts for the algebraic
widening of the front. 

Far from the front region, we anticipate the following extremal behavior
of the scaling function $F(z)$:
\begin{equation}
\label{extremal}
F(z)\sim\cases{1-\exp(-z^{\sigma_+})&$z\to\infty$;\cr
               \exp(-|z|^{\sigma_-})&$z\to-\infty$.}
\end{equation}
The exponents $\sigma_{\pm}$ thus characterize relaxation away from the
front region. The exponent $\sigma_+$ which describes large positive
fluctuations in the height can be simply related to the roughness
exponent.  Consider a large positive height fluctuation, $j=Avt$, with
$A\gg 1$.  Such large ``towers'' can be created only by an anomalously
large number, $Avt$, of nucleation events localized in the same region.
Given the Poisson nature of the nucleation events, such fluctuations are
suppressed exponentially.  Thus, the quantity $1-S_{Avt}(t)$ is
estimated by $\exp(-t)$, but since Eq.~(\ref{extremal}) gives
$\exp\left[-t^{\sigma_+(1-\beta)}\right]$, we conclude that

\begin{equation}
\label{sigp}
\sigma_+={1\over 1-\beta}.
\end{equation}

\section{Mean Field Theory}

In the following, we explore the {\em non-equilibrium} regime, i.e.,
we consider polynuclear growth on an infinite substrate.
Non-equilibrium behavior should agree with the early time behavior on
finite substrates, since then finite size effects are still
negligible.

\subsection{One Dimension}

We start with the PNG model in one dimension where a more comprehensive
analysis is possible.  In this situation, steps nucleate in pairs and
move away from each other with a constant velocity.  The constant
nucleation rate and the growth velocity are set to unity, without loss
of generality.  Hence, the length of an island at time $t$ after birth
equals $2t$.  Consider $f_j(x,t)$, the density of gaps of length $x$ at
time $t$ in the $j$th layer.  This distribution evolves according to
\begin{eqnarray}
\label{fjt}
{\partial f_j(x,t)\over \partial t}&=&2{\partial f_j(x,t)\over \partial x}\\
&+&\gamma_j(t)\left[-xf_j(x,t)+2\int_x^\infty dy f_j(y,t)\right].
\nonumber
\end{eqnarray}
The spatial derivative term describes shrinkage of gaps.  The last two
terms account for changes due to nucleation and thus, are proportional
to the overall nucleation rate at the $j$th layer, $\gamma_j(t)$. The
loss term is proportional to the gap length and the gain term describes
creation of gaps from larger gaps.

Eqs.~(\ref{fjt}) contain yet unknown nucleation rates $\gamma_j(t)$
which will be chosen to satisfy the correct kinetic equations for the
uncovered fractions
\begin{equation}
\label{S}
S_j(t)=\int_0^\infty dx\, x\, f_j(x,t),
\end{equation}
and the gap (or island) densities
\begin{equation}
\label{N}
N_j(t)=\int_0^\infty dx f_j(x,t).
\end{equation}
The uncovered fraction decreases with a rate proportional to the island
density, $\dot S_j(t)=-2N_j(t)$, and by integrating Eqs.~(\ref{fjt}) we
indeed recover this exact equation. The island density changes due to
disappearance of gaps as well as due to nucleation. The total nucleation
rate is proportional to the exposed fraction of the $j$th layer and
thus, $\dot N_j(t)=-2f_j(0,t)+S_j(t)-S_{j-1}(t)$.  On the other hand,
by integrating  Eqs.~(\ref{fjt}) we obtain $\dot N_j(t)=
-2f_j(0,t)+\gamma_j(t)S_j(t)$.  Therefore, the choice
\begin{equation}
\label{gamma}
\gamma_j(t)=1-{S_{j-1}(t)\over S_j(t)}
\end{equation}
guarantees that the gap density evolves according to {\em exact} rate
equation.  The expression of Eq.~(\ref{gamma}) for the nucleation rate
$\gamma_j(t)$ in the $j$th layer is intuitively appealing since the
total unit nucleation rate should be reduced to account for nucleation
events below the $j$th layer.  Taking into account that $S_0(t)\equiv 0$
we find $\gamma_1(t)\equiv 1$ and we notice that Eq.~(\ref{fjt}) for the
first layer agrees with  the exact KAJM equation\cite{sek}.  Thus, the set
of rate equations (\ref{fjt}) with the nucleation rates (\ref{gamma})
provides a self-consistent description of the 1d PNG model. It is exact
for the first layer, and additionally, the first two moments of the gap
density satisfy the correct rate equations.  However, it is a mean-field
description since it assumes a spatially homogeneous nucleation rate
$\gamma_j(t)$.

The gap density is found to be exponential, and the formal solution reads
\begin{equation}
\label{fgjt}
f_j(x,t)=g^2_j(t)\,\exp\left[-g_j(t)x-2\int_0^t d\tau
  g_j(\tau)\right], 
\end{equation}
with $g_j(t)=\int_0^t d\tau\,\gamma_j(\tau)$.  The uncovered fraction
and the island density are evaluated using Eqs.~ (\ref{S}) and (\ref{N}):
\begin{eqnarray}
S_j(t)&=&e^{-2\int_0^t d\tau g_j(\tau)},\nonumber\\
N_j(t)&=&g_j(t)S_j(t).
\end{eqnarray}
Evaluating $d^2\ln S_j(t)/dt^2$ together with Eq.~(\ref{gamma}) and $\dot
g_j(t)=\gamma_j(t)$, leads to an infinite set of recursive 
differential equations for the uncovered fraction
\begin{equation}
\label{sj}
\ddot S_j-\dot S_j^2S_j^{-1}+2(S_j-S_{j-1})=0.
\end{equation}
Eqs.~(\ref{sj}) should be solved subject to the initial conditions
$S_j(0)=1$ and $ \dot S_j(0)=0$ for $j\geq 1$.  The recursive
structure of Eqs.~(\ref{sj}) reflects the fact that kinetics of a
given layer is {\em unrelated} to that of all layers above (thanks to
the absence of overhangs).  Additionally, Eqs.~(\ref{sj}) agree with
the nature of the PNG model implying that kinetics of a given layer
should be {\em directly} coupled only to the previous layer.

Using $S_0(t)\equiv 0$, $S_1(t)$ is determined, then $S_2(t)$, etc.  Of
course, for the first layer, the KAJM nucleation-and-growth results are
reproduced \cite{sek}
\begin{eqnarray}
\label{s1}
S_1(t)&=&e^{-t^2},\\
f_1(x,t)&=&t^2e^{-xt-t^2}\nonumber. 
\end{eqnarray}
It is also possible to solve analytically for the second layer 
\begin{eqnarray}
\label{s2}
S_2(t)&=&\cosh^2 t\,e^{-t^2},\\
f_2(x,t)&=&[t\,\cosh\,t-\sinh\,t]^2\,
e^{-(t-\tanh\,t)x-t^2}. \nonumber
\end{eqnarray}
Using the transformation, $S_j(t)=\exp\left[u_j(t)-t^2\right]$, the
differential equations (\ref{sj}) formally simplify to a directed
version of the Toda equations \cite{toda}, $\ddot
u_j=2\exp[u_{j-1}-u_j]$, with the initial conditions $u_j(0)=\dot
u_j(0)=0$ and the boundary condition $u_1(t)=0$. Despite this
simplification it is not possible to integrate these equations, and we
solve numerically for $S_j(t)$.  Fig.~1 shows how the coverage in a
given layer changes with time.  We see that the coverage quickly relaxes
onto a traveling wave form with a finite width, $S_j(t)\to F(j-vt)$.

Some quantitative characteristics of the traveling wave can be
determined analytically.  For $j-vt\gg 1$, the nonlinear term in
(\ref{sj}) is negligible and Eqs.~(\ref{sj}) become linear.  Thus, we
write
\begin{equation}
\label{ansatz}
1-S_j(t)\sim e^{-\alpha(j-vt)}, \quad j-vt\gg 1, 
\end{equation}
with a yet unknown coefficient $\alpha$.  Substituting  into
Eqs.~(\ref{sj}) gives
\begin{equation}
\label{kol}
v^2=2{e^{\alpha}-1\over \alpha^2}.
\end{equation}
The right hand side has a minimum at $\alpha=1.59362$.  Therefore any
velocity in the interval $[v_{\rm min},\infty)$ with $v_{\rm min}
=1.75735$ is possible.  Our numerical integration shows a velocity
that falls within 0.1\% of $v_{\rm min}$, thereby implying that the
minimal velocity is indeed selected.  Such minimum velocity selection
is ubiquitous and occurs for a wide class of initial 
conditions\cite{br,mur}.

\begin{figure}
\vspace{-.2in}
\centerline{\epsfxsize=9cm \epsfbox{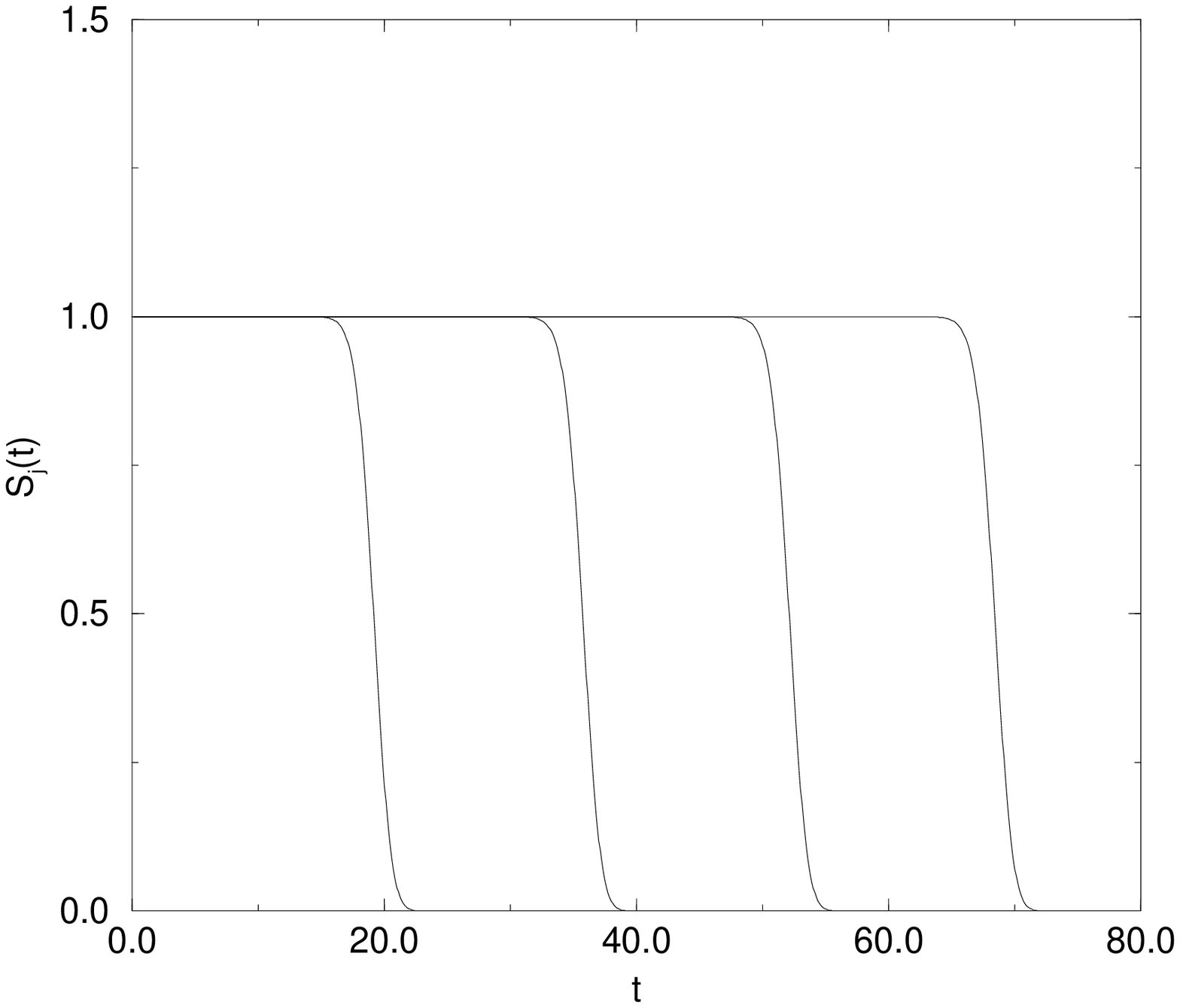}}
\vspace{-.2in}
\noindent{\small {\bf Fig.1}
  The uncovered fraction $S_j(t)$ vs. time for layers $j=$ 20, 40, 60,
  and 80.  Clearly, the coverage follows a traveling wave form.}
\end{figure}

To obtain the asymptotic behavior in the other extreme, $z=j-vt\to
-\infty$, we first note that $S_2(t)\gg S_1(t)$, as follows from
(\ref{s1}) and (\ref{s2}).  We assume this for all layers far behind the
front; we will check this assumption $S_j(t)\gg S_{j-1}(t)$ a
posteriori.  This reduces Eqs.~(\ref{sj}) to $\ddot S_j-\dot
S_j^2S_j^{-1}+2S_j=0$ which is solved to yield
$S_j(t)=\exp(-t^2+A_jt+B_j)$.  The traveling wave form implies that
$S_j(t)$ should be a function of a single variable $z=j-vt$.  This
determines the constants $A_j$ and $B_j$, and we find $S_j(t)=F(z)\sim
\exp\left(-z^2/v^2\right)$.  One can verify that the assumption
$S_j(t)\gg S_{j-1}(t)$ is valid.

\subsection{Arbitrary Dimension}

The above analysis cannot be generalized in a straightforward manner
to $d\ne 1$ since the gap distribution is intrinsically
one-dimensional. However, it is still possible to obtain a mean field
description for the uncovered fraction.  

Consider first a KAJM nucleation-and-growth process where the rate of
nucleation events $\gamma(t)$ is homogeneous in space but time
dependent.  Ignoring overlap between growing disks, the uncovered
fraction $S(t)$ decreases with time according to
\begin{equation}
\label{no}
{dS(t)\over dt}=-d\Omega_d\int_0^t d\tau \gamma(\tau) (t-\tau)^{d-1},
\end{equation}
where $\Omega_d=\pi^{d/2}/\Gamma(1+d/2)$ is the volume of the
$d$-dimensional unit sphere and $d\Omega_d$ is its surface area
\cite{Bender} ($\Gamma$ is the gamma function).  Of course,
Eq.~(\ref{no}) overestimates the decay rate since some of the area is
already covered.  Nevertheless, this may be corrected if $(-dS/dt)$ is
reduced by the uncovered fraction, $S$, 
\begin{equation}
\label{yes}
{dS(t)\over dt}=-d\Omega_d\,S(t)\,\int_0^t\, d\tau\,\gamma(\tau)
(t-\tau)^{d-1}.
\end{equation}
For a constant nucleation rate, $\gamma=1$, Eq.~(\ref{yes}) gives 
\begin{equation}
\label{first}
S(t)=\exp\left[-{\Omega_d t^{d+1}\over d+1}\right].
\end{equation}
Thus, the exact KAJM coverage is recovered.  Furthermore, a
generalization of the KAJM solution to time-dependent nucleation rates
appears to be equivalent to Eq.~(\ref{yes}) \cite{Sek}.

We now return to the PNG model. Note first that the nonlocal in time
integro-differential Eq.~(\ref{yes}) can be converted to a higher
order ordinary differential equation $d^{d+1}\ln
S_j(t)/dt^{d+1}=-d!\,\Omega_d \gamma_j(t)$.  Unlike the 1d case, it is
not possible to derive the nucleation rate self-consistently. However,
assuming a spatially homogeneous nucleation rate \cite{gilmer} implies
the total nucleation rate $\gamma_j(t)S_j(t)=S_j(t)-S_{j-1}(t)$ and
therefore Eq.~(\ref{gamma}).  We thus arrive at the following
generalization of the mean-field Eq.~(\ref{sj}) for the uncovered
fraction
\begin{equation}
\label{dsj}
{d^{d+1} \over dt^{d+1}}\,\ln S_j+d!\,\Omega_d
\left[1-{S_{j-1}(t)\over S_j(t)}\right]=0.
\end{equation}

The analysis presented in the one-dimensional case applies for
arbitrary dimensions. For example, the transformation
$S_j(t)=\exp\left[u_j(t)-\Omega_d t^{d+1}/(d+1)\right]$ reduces
Eqs.~(\ref{dsj}) to a set of generalized directed Toda equations
$d^{d+1} u_j/ dt^{d+1}=d!\Omega_d\exp[u_{j-1}-u_j]$.  Analysis of
these equations or Eqs.~(\ref{dsj}) reveals that the coverage relaxes 
to a traveling wave with a finite width.  To determine the growth
velocity, we insert the ansatz of Eq.~(\ref{ansatz}) into
Eqs.~(\ref{dsj}) to find
\begin{equation}
\label{dv}
v^{d+1}=d!\Omega_d\,{e^{\alpha}-1\over \alpha^{d+1}}.
\end{equation}
This provides the lower bound for the growth velocity, $v_d\geq
v_d^{\rm min}$.  Again the minimal velocity should be selected, and
for example $v_0=1$, $v_1=1.75735$, $v_2=1.67115$, and $v_d\simeq
\sqrt{2\pi e/d}$ when $d\to\infty$.  Furthermore, in finite dimensions
mean field theory predicts universal exponents $\beta=0$,
$\sigma_+=1$, and $\sigma_-=2$. (Note that the relationship of 
Eq.~(\ref{sigp}) is obeyed).

In the zero-dimension limit, the behavior changes qualitatively.
Indeed, for the $d=0$ case Eqs.~(\ref{dsj}) become {\em linear},
\begin{equation}
\label{lin}
{d S_j\over dt}+S_j=S_{j-1}.  
\end{equation}
Solving (\ref{lin}) recursively yields the uncovered fraction,
$S_j(t)=e^{-t}\sum_{i=0}^{j-1} t^i/i!$.  Alternatively, by treating the
variable $j$ as continuous, this set of linear equations reduces to a
simple convection-diffusion equation with a unit velocity and a
diffusion coefficient $D=1/2$. Consequently, the roughness becomes
diffusive, i.e., $\beta=1/2$.

\section{LINEAR RECURSION RELATIONS}

The Linear Recursion Relations (LRR) approach employs the exact
uncovered fraction $S_1(t)$ in the first layer, provided by the KAJM
solution (\ref{first}) \cite{kas,be,ah,ran}.  Nucleation in the
$(j+1)$st layer proceeds only on the already covered fraction of the
$j$th layer, formed with rate $-dS_j/dt$.  Subsequent covering
proceeds as in the KAJM, so one anticipates that nucleation events in
the time interval $(\tau,\tau+d\tau)$ make a contribution
$S_1(t-\tau)\left[-\dot S_j(\tau)\right] d\tau$ to the exposed
fraction $S_{j+1}(t)-S_j(t)$ in the $(j+1)$st layer.  This leads to a
recursion relation between adjacent layers:
\begin{equation}
\label{next}
S_{j+1}(t)=S_j(t)-\int_0^t d\tau\, S_1(t-\tau)\,{dS_j(\tau)\over d\tau}.
\label{tj}
\end{equation}
The first layer coverage (\ref{first}) can be recovered by setting the
substrate coverage appropriately, $dS_0/dt=-\delta(t)$.  

Although MFT and LRR are both recursive as every layer is coupled to
the preceding layer, they differ in that the MFT equations are
nonlinear, while the LRR equations are linear. Nevertheless, when
$d\to 0$, both approximations are identical. Indeed, multiplying
Eq.~(\ref{next}) by $e^t$ and differentiating, one recovers the MFT
equation (\ref{lin}).

Using the Laplace transform, analytical results for the growth
velocity and the interface width have been established
\cite{ran,be,new}.  Below, we give an alternative and simpler
derivation which additionally provides the asymptotic behavior of the
coverage profile.  In the long time limit, it is reasonable to treat
the layer number $j$ as a continuous variable.  Replacing the
difference $S_{j+1}-S_j$ by a partial derivative, the recursive
relations (\ref{tj}) become
\begin{eqnarray}
\label{cov}
{\partial S\over \partial j}&\cong& -\int_0^t d\tau\,S_1(\tau)\,
{\partial S(j,t')\over \partial t'}\bigg|_{t'=t-\tau}\\
&\cong& -{1\over v_d}{\partial S\over \partial t}
+{1\over 2v_d^2}\,{\Gamma\left({d+3\over d+1}\right)\over
\Gamma^2\left({d+2\over d+1}\right)}
{\partial^2 S\over \partial t^2},\nonumber
\end{eqnarray}
with $v_d$ the growth velocity 
\begin{equation}
\label{vdim}
v_d=\left({\Omega_d\over d+1}\right)^{1\over d+1}
\Big /\Gamma\left({d+2\over d+1}\right).
\end{equation}
The second line in (\ref{cov}) has been derived by expanding
$S(j,t-\tau)$ in a Taylor series in $\tau$, keeping only the two
dominant terms of the expansion, and replacing the upper limit in the
integral by $\infty$.  The following growth velocities are found:
$v_0=1$, $v_1=2/\sqrt{\pi}=1.12838$, $v_2=1.13719$, and $v_d\to
\sqrt{2\pi e/d}$ as $d\to\infty$. The velocity is almost constant for
physical dimensions (it varies by less than 4\% in the range $1\le d \le
4$) indicating the weak dimension dependence of this approach.

Changing variables from $(j,t)$ to $(j, \xi=j-v_dt)$ recasts
Eq.~(\ref{cov}) into a diffusion equation
\begin{equation}
{\partial S\over\partial t}=D{\partial^2 S\over\partial \xi^2}. 
\label{diff}
\end{equation}
The constant $D={v_d\over 2}\,\Gamma\left({d+3\over d+1}\right)\Big/
\Gamma^2\left({d+2\over d+1}\right)$ plays the role of a diffusion
coefficient and  controls the width of the interface.  In
obtaining Eq.~(\ref{diff}), $j$ was replaced by $v_dt$.  This is clearly
valid in the scaling limit, $j\to\infty$, $|\xi|\to\infty$, $j\sim \xi^2$.
The initial profile of the uncovered fraction, $S(j,0)$, is a step
function: $S(j,0)=0$ for $j\leq 0$ and $S(j,0)=1$ for $j>0$.  Solving
(\ref{diff}) subject to these initial conditions yields
\begin{equation}
S(j,t)={1\over 2}{\rm Erfc}(-z), \quad 
z={\xi\over\sqrt{4Dt}}={j-v_dt\over \sqrt{4Dt}},
\label{cover}
\end{equation} 
with ${\rm Erfc}(z)={2\over \sqrt{\pi}}\int_z^{\infty} du\, e^{-u^2}$
the error function\cite{Bender}. While the two approximations
generally differ in their asymptotic behavior, they do agree in the
extreme cases of $d=0$ and $d=\infty$.

In summary, the LRR approach predicts a dimension-independent
``diffusive'' width exponent $\beta=1/2$.  The shape of the coverage
profile is symmetric and Gaussian far from the front,
$\sigma_{\pm}=2$.

\section{Comparison with Simulations}

To test the two approximations we simulated the PNG process. The
simulation results presented below are for a one-dimensional chain of
length $L=10^4$ and represent a single realization. This study is
different than previous numerical studies which simulated the PNG
process on a lattice\cite{kr}.  Here we treated time and space as
continuous variables, and this enables comparison with the above
theories.

The time dependence of the uncovered fraction for the first four layers
is shown in Fig.~2.  It is seen that the MFT and the LRR approaches
provide upper and lower bounds, respectively, for the actual PNG
coverage.  Additionally, the MFT provides a better approximation for the
uncovered fraction, $S_j(t)$.  For early times, the height and width
predicted by either approximations are quite close to simulation
results, as shown in Fig.~3.  In fact, both approaches agree to the
first significant order in time, as both Eq.~(\ref{sj}) and
Eq.~(\ref{next}) predict $S_j(t)=1-{2^j\over (2j)!}t^{2j}$. The
disagreement between the two is of the order $t^{2j+2}$. As the two
approximations give upper and lower bounds for the PNG process, we
conclude that this is the leading early time behavior of $S_j(t)$.

\begin{figure}
\vspace{-.24in} 
\centerline{\epsfxsize=9cm \epsfbox{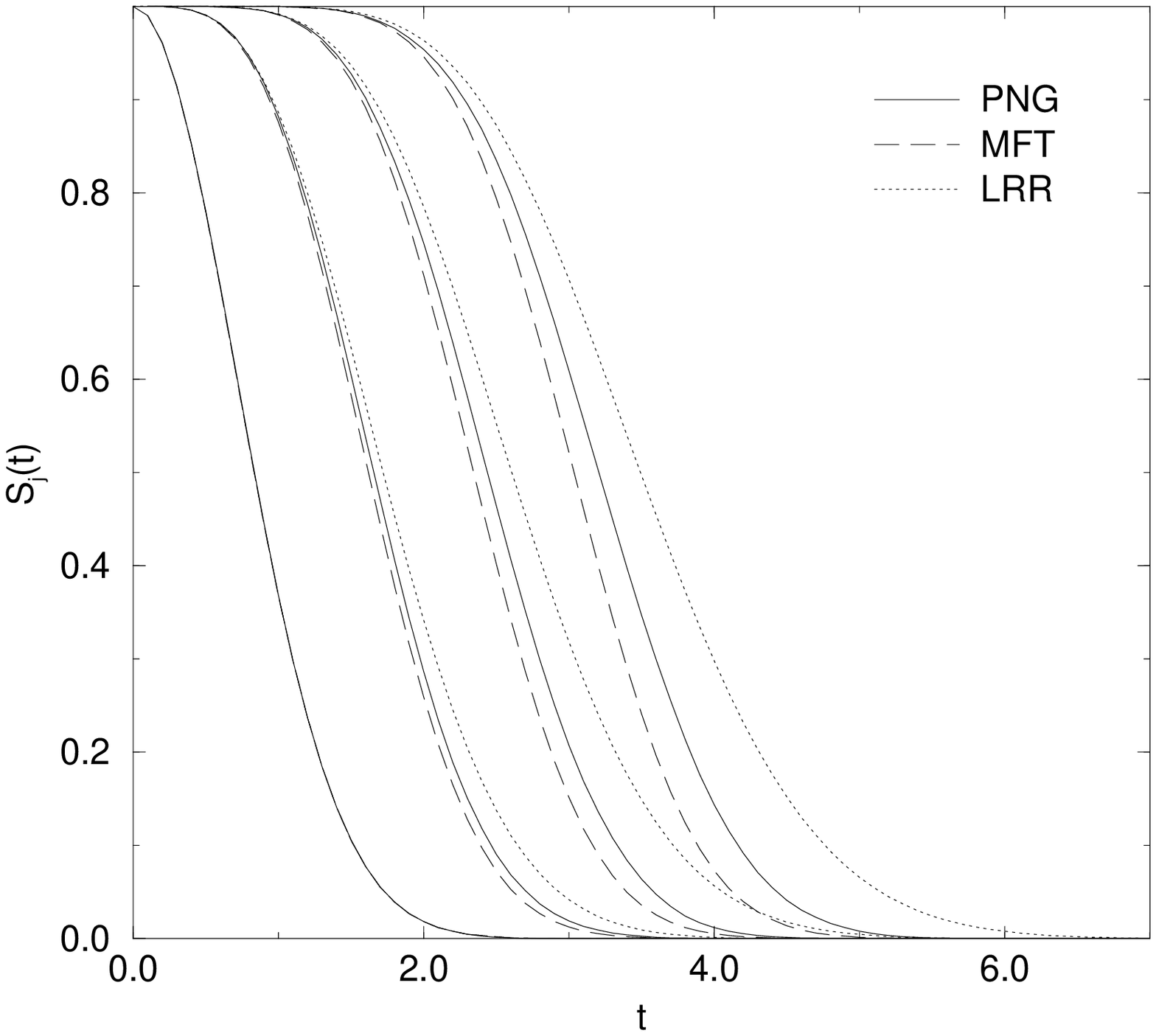}}
\vspace{-.35in} 
\noindent{\small {\bf Fig.2.}  
Uncovered fraction $S_j(t)$ versus $t$ for $j=1,2,3,4$. MFT and
LRR approaches give lower and upper bounds, respectively, for the 
uncovered fraction.}
\end{figure}

\begin{figure}
\vspace{-.24in}
\centerline{\epsfxsize=9cm \epsfbox{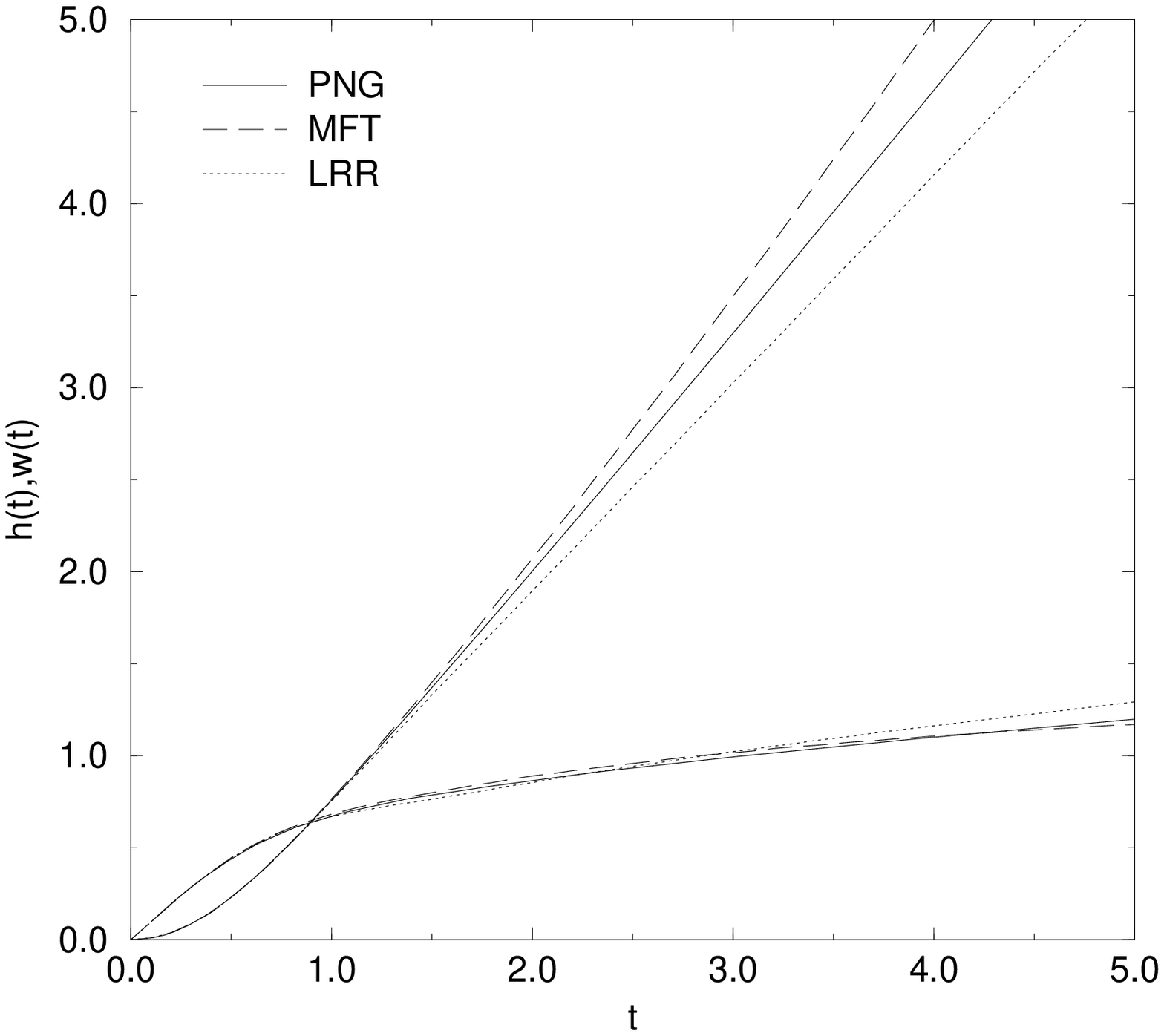}}
\vspace{-.35in}
\noindent{\small {\bf Fig.3.}
Short time behavior of the height $h(t)$ and the width $w(t)$. 
MFT is closer to the actual behavior.}
\end{figure}

However, both approximations become progressively worse at later
times. This is due to the fact that the asymptotic behavior of the
width is predicted incorrectly (see Table 1). Fig.~4 shows that at
least in one dimension, the PNG asymptotic behavior belongs to the KPZ
universality class. As our simulations are continuous in space and
time, they enable measurement of the surface growth velocity.
Although in principle, there is no reason to expect that the surface
growth velocity in an infinite and a finite system are equal, the
numerical velocity $v_{\rm noneq}=1.41\pm 0.01$, which corresponds to
(non-equilibrium) growth on an infinite substrate, is in very good
agreement with the known analytical equilibrium growth velocity
$v_{\rm eq}=\sqrt{2}$\cite{ben}.  Numerical values for $v$, $\beta$
and $\sigma_{\pm}$ are summarized in Table 1. Two-dimensional
simulations give a velocity $v_2\approx 1.4$ \cite{gilmer} compare to
the values $v_2=1.671$ (MFT) and $v_2=1.137$ (LRR).  This suggests
that MFT improves for higher dimensions.  Furthermore, the exponent
$\beta$ decreases with the dimension and it vanishes for $d\ge
4$\cite{zhang}. Above this critical dimension, mean-field behavior
occurs.

We also examined the extremal behavior of $S_j(t)$ using the
simulations. The scaling prediction (\ref{sigp}) holds as the
simulation data is consistent with the exponent value $\sigma_+=3/2$.
Given that the PNG uncovered fraction is bounded by the two
approximations, the parameters $v$, $\beta$, and therefore $\sigma_+$
are similarly bounded (see Table 1.)  Since the scaling argument
involves the width, one can not conclude a priory that the same holds
for $\sigma_-$.  Nevertheless, the exponent found in the simulation is
quite close to $2$, or possibly slightly larger.

\begin{figure}
\vspace{-.35in}
\centerline{\epsfxsize=9cm \epsfbox{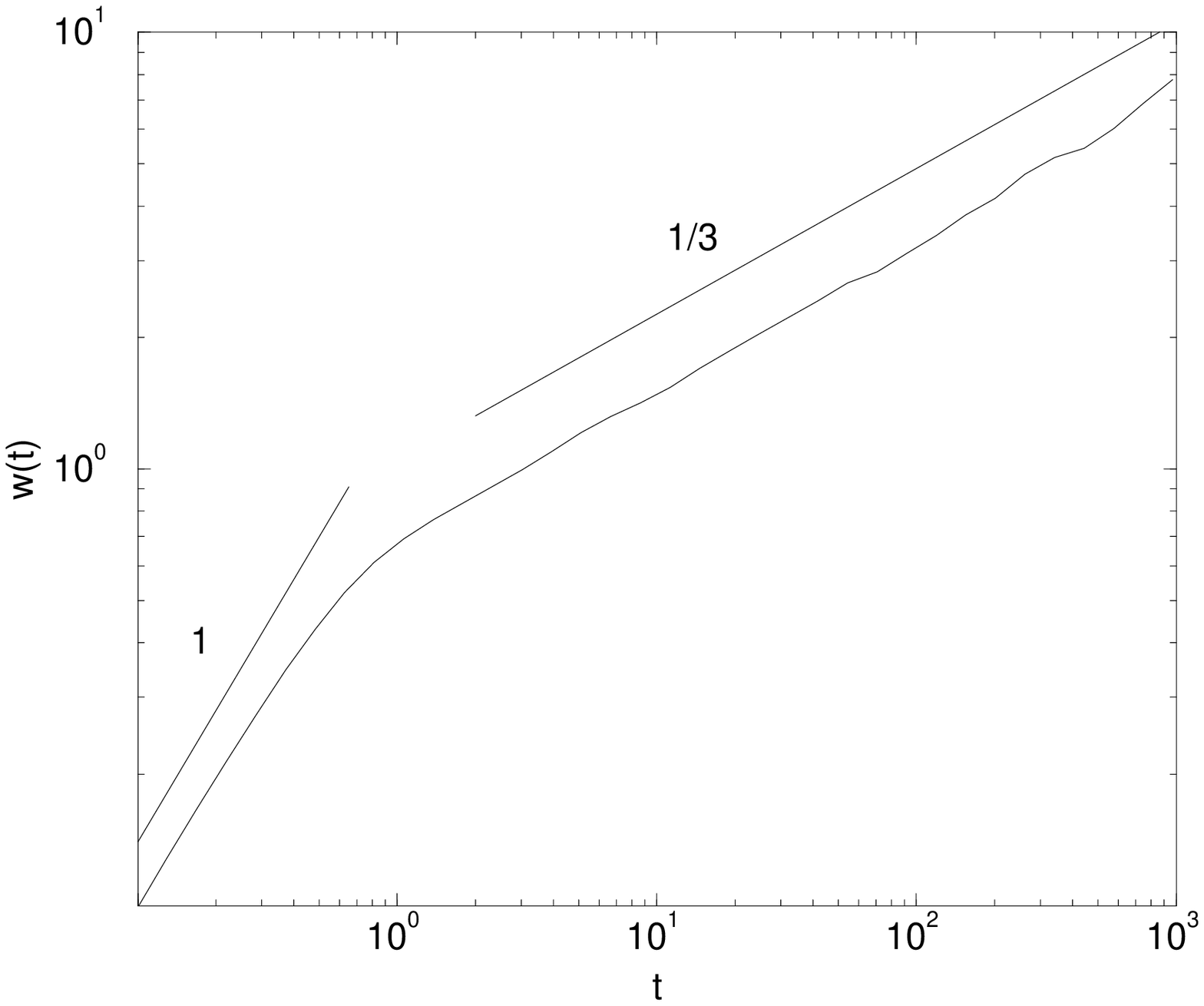}}
\vspace{-.25in}
\noindent{\small {\bf Fig.4}
Long time behavior of the width. Early behavior is linear and 
late behavior is in the KPZ universality class $t^{1/3}$.}
\end{figure}

\vspace{.1in}

\centerline{
\begin{tabular}{|c|c|c|c|}
\hline &MFT&PNG&LRR\\ 
\hline
$v_1$&1.75735&$1.41\pm0.01$&1.12838\\
$\beta$&0&1/3&1/2\\
$\sigma_+$&1&3/2&2\\
$\sigma_-$&2&$\ge 2$&2\\
\hline
\end{tabular}}
\vspace{.1in}
\noindent{\small {\bf Table 1}: Characteristics of the three 
approaches for the one-dimensional PNG model.}

\section{CONCLUSIONS}
 
We have investigated a continuum model of multilayer growth, the PNG
model.  We confirmed that the LRR approximation implies a
dimension-independent asymptotic behavior with the roughness exponent
$\beta=1/2$.  Moreover, we found that the full coverage profile
approaches a dimension independent form, thus implying that this
approximation ignores interactions.  We developed an alternative
self-consistent mean field approach that provides a better
approximation for the uncovered fraction, and close estimates for the
early time behavior.  Additionally, the two approximate approaches
combine to give upper and lower bounds for statistical properties such
as the coverage, the velocity, and the roughness.

The mean field approach predicts a smooth interface, $\beta=0$.  With
increasing the dimension, the roughness exponent decreases from
$\beta=1/2$ at $d=0$ to $\beta=0$ at $d\ge d_c$, with $d_c$ the
critical dimension\cite{zhang}.  Thus, for sufficiently high dimensions
the mean-field behavior should emerge.

The above mean-field theory should allow computation of space-time
correlation functions and structure functions.  Even more detailed
analysis may be possible for one-dimensional substrates.  A bigger
challenge is to solve the PNG process analytically.  Such a solution
will undoubtedly illuminate the theoretical understanding of
non-equilibrium growth.

\vskip 0.1in
\noindent
PLK acknowledges support from NSF and ARO.

\end{multicols}
\end{document}